# Set graphs. II. Complexity of set graph recognition and similar problems

Martin Milanič[a,b], Romeo Rizzi[c], Alexandru I. Tomescu[c,d,*]

[a]*Faculty of Mathematics, Natural Sciences and Information Technologies*
*University of Primorska, Glagoljaška 8, 6000 Koper, Slovenia*
[b]*Institute Andrej Marušič*
*University of Primorska, Muzejski trg 2, 6000 Koper, Slovenia*
[c]*Dipartimento di Matematica e Informatica, Università di Udine,*
*Via delle Scienze, 206, 33100 Udine, Italy*
[d]*Faculty of Mathematics and Computer Science, University of Bucharest,*
*Str. Academiei, 14, 010014 Bucharest, Romania*

**Abstract**

A graph $G$ is said to be a *set graph* if it admits an acyclic orientation that is also *extensional*, in the sense that the out-neighborhoods of its vertices are pairwise distinct. Equivalently, a set graph is the underlying graph of the digraph representation of a hereditarily finite set. In this paper, we continue the study of set graphs and related topics, focusing on computational complexity aspects. We prove that set graph recognition is NP-complete, even when the input is restricted to bipartite graphs with exactly two leaves. The problem remains NP-complete if, in addition, we require that the extensional acyclic orientation be also *slim*, that is, that the digraph obtained by removing any arc from it is not extensional. We also show that the counting variants of the above problems are #P-complete, and prove similar complexity results for problems related to a generalization of extensional acyclic digraphs, the so-called *hyper-extensional digraphs*, which were proposed by Aczel to describe hypersets. Our proofs are based on reductions from variants of the HAMILTONIAN PATH problem. We also consider a variant of the well-known notion of a separating code in a digraph, the so-called *open-out-separating code*, and show that it is NP-complete to determine whether an input extensional acyclic digraph contains an open-out-separating code of given size.

## 1. Introduction

The view of a set as a digraph is almost as old as sets themselves. Even though for representing graphs one uses sets, as *flat* collections of vertices, or edges, digraphs enter into play in capturing the *nested* membership structure of a pure set. For this, one must consider the collection of its elements, of the elements of its elements, and so on. This leads to the notion of the *transitive closure* of a set $x$, defined as $\mathsf{TrCl}(x) = x \cup \bigcup_{y \in x} \mathsf{TrCl}(y)$. The *membership digraph* associated to

$x$ has $\mathsf{TrCl}(x)$ as the vertex set and the inverse of the membership relation as the arc relation:

$$(\mathsf{TrCl}(x), \{u \to v \mid u, v \in \mathsf{TrCl}(x), v \in u\}).$$

Clearly, such a digraph is acyclic—as $\in$ is well-founded—, and *extensional*, in the sense that different vertices have different sets of out-neighbors. It can be easily seen that there actually exists a bijection between membership digraphs and extensional acyclic digraphs (*e.a.* digraphs, for short). This is given by the so-called *Mostowski's collapse* of an e.a. digraph, which recursively associates to each vertex the set of sets associated to its out-neighbors (cf., e.g., [1]).

With the goal of pursuing a systematic study of the *topological* diversity of the underlying (undirected) graphs of extensional acyclic digraphs, Milanič and Tomescu introduced in [2] the notion of a *set graph*, that is, a graph that admits an e.a. orientation. Not every graph is a set graph—for example, the *claw*, $K_{1,3}$, is not—but, remarkably, all connected claw-free graphs are set graphs, and so are all graphs containing a Hamiltonian path [2]. In the same paper [2], several necessary conditions for a graph to be a set graph were established, unicyclic set graphs were characterized, a polynomial time algorithm was given for finding an extensional acyclic orientation of a connected claw-free graph, and two problems closely related to the problem of recognizing set graphs were shown to be NP-complete in general but—together with set graph recognition—proved to be solvable in linear time on graphs of bounded treewidth.

Besides being useful for representing hereditarily finite sets—that is, sets the transitive closure of which is finite—, extensional acyclic digraphs, and therefore also set graphs, find further applications. The study of extensional acyclic digraphs inspired short proofs of two classical results on claw-free graphs (a strengthening of the fact that squares of connected claw-free graphs are Hamiltonian, and the fact that every connected claw-free graph of even order has a perfect matching) [2]. A formalization of these two results in the proof-checker Referee was carried out in [3]. Since Referee deals only with Zermelo-Fraenkel sets, representing a connected claw-free graph by a transitive 'claw-free' set turned out to require the minimal formalism. Many other results on set graphs can probably be formalized in this way in Referee. On the more practical side, extensional acyclic digraphs find applications, e.g., in the design of emergency sensor networks in facilities or in fault detection in multiprocessor systems. See [2] for a more detailed discussion and further references.

In this paper, we continue the study of set graphs, focusing on computational complexity aspects. We consider three problems closely related to set graph recognition, and establish NP-completeness for each of them:

- We prove that set graph recognition is NP-complete, even when the input is restricted to bipartite graphs with exactly two leaves.

- We prove that it is NP-complete to determine whether the input graph admits an extensional acyclic orientation that is also *slim*, that is, the digraph obtained by removing any arc from it is not extensional. The same result holds if the input graph is a bipartite graph with exactly two leaves.

- We consider a generalization of extensional acyclic digraphs, the so-called *hyper-extensional digraphs*, which were proposed by Aczel [4] to describe hypersets. We prove that it is NP-complete to determine whether the input graph admits a hyper-extensional orientation, even if the input graph is a bipartite graph with exactly three leaves.

Our proofs are based on reductions from variants of the HAMILTONIAN PATH problem. We also prove that the counting variants of the three above problems are #P-complete. Finally, we consider a variant of the well-known notion of a separating code in a digraph, the so-called *open-out-separating code*, and show that it is NP-complete to determine whether an input extensional acyclic digraph contains an open-out-separating code of given size.

The paper is structured as follows. In Section 2, we establish the notation and necessary preliminary results. In Section 3, we state and prove the NP-completeness results of the decision problems mentioned above. In Section 4 is devoted to counting variants of the three decision problems and proofs of the corresponding #P-completeness results. In Section 5 we prove the NP-completeness result related to open-out-separating codes. We conclude the paper with some open questions in Section 6.

## 2. Notation and preliminaries

We consider finite *simple* graphs and digraphs, that is, (di)graphs without loops or parallel edges/arcs. Given a digraph $D$, we denote an arc $(x,y) \in E(D)$ as $xy$, or $x \to y$. Moreover, for any $x \in V(D)$ we denote by $N_D^+(x)$ the set of out-neighbors of $x$ in $D$, i.e., the set $\{y \in V(D) \mid xy \in E(D)\}$. We may skip the subscript $D$ when it is clear from the context. Similarly, $N_G(x)$ is the set of neighbors of $x$ in a graph $G$. A vertex $x$ in a digraph $D$ is a *sink* if it has no out-neighbors, and a *source* if it has no in-neighbors.

Given a (di)graph $G$ and $S \subseteq V(G)$, we denote by $G - S$ the vertex-induced sub(di)graph $G[V(G) \setminus S]$. When $S = \{x\}$, we will write $G - x$ instead of $G - \{x\}$. A vertex $v$ of a graph $G$ is *cut vertex* if $G - x$ has more connected components than $G$. As usual, we denote by $K_{m,n}$ the complete bipartite graph with parts of size $m$ and $n$. The *claw* is the graph $K_{1,3}$. For two graphs $G$ and $H$, we say that $G$ is $H$-free if no induced subgraph of $G$ is isomorphic to $H$. In particular, a graph with no induced claw is said to be *claw-free*. Given a (di)graph $G$ and an induced sub(di)graph $H$ of $G$, we may write $H$ instead of $V(H)$, whenever this is clear from the context. The *rank* of a vertex $v$ in an acyclic digraph is defined as the length of a longest directed path from $v$ to a sink of $D$.

An *orientation* of a graph $G = (V, E)$ is a digraph $D = (V, E')$ such that $|E'| = |E|$ and $uv \in E$ for every $(u, v) \in E'$. Graph $G$ is called the *underlying graph* of $D$.

**Definition 1.** *A digraph $D$ is said to be* extensional *if for every two distinct vertices $u$ and $v$ in $V(D)$, it holds that $N^+(v) \neq N^+(u)$.*

We say that a digraph $D$ is an extensional acyclic orientation—e.a.o., for short—of a graph $G$ if $G$ is the underlying graph of $D$ and $D$ is acyclic and extensional. Whenever in a digraph $D$ for distinct vertices $x$ and $y$ we have $N^+(x) = N^+(y)$, we say that $x$ and $y$ *collide*. Note that this is not the case if $D$ is acyclic and there is a directed path from $x$ to $y$.

**Definition 2** ([2]). *A graph $G$ is said to be a* set graph *if $G$ admits an extensional acyclic orientation (e.a.o., for short).*

The following refinement of extensionality was introduced in [5], where it was employed to identify a class of extensional acyclic digraphs well-quasi-ordered by the strong immersion relation.

**Definition 3.** *An extensional acyclic digraph $D$ is said to be* slim *if the digraph obtained by removing any arc from $D$ is not extensional.*

The following is a broad-range notion appearing in many fields of theoretical computer science: modal logic [6], concurrency theory [7, 8], formal verification [9].

**Definition 4.** A bisimulation *over a digraph D is a relation* $B \subseteq V(D) \times V(D)$ *such that* $xBy$ *implies that*

i) *for every* $x'$ *such that* $x \to x'$ *holds, there exists a vertex* $y'$ *such that* $y \to y'$ *and* $x'By'$; *and*

ii) *for every* $y'$ *such that* $y \to y'$ *holds, there exists a vertex* $x'$ *such that* $x \to x'$ *and* $x'By'$.

Usually, given a digraph $D$, one is interested in deciding whether for two vertices $x, y$ of $D$ there exists a bisimulation relation $B$ over $D$ such that $xBy$ holds. To answer this problem in an efficient way, it is convenient to compute the *maximum bisimulation* over $D$, that is, the equivalence relation which is the union of all bisimulation relations over $D$. This problem was shown in [10] to be equivalent to the stable partitioning problem, which is solvable in time $O(|E(D)|\log|V(D)|)$ by the algorithm [11] of Paige and Tarjan. The complexity of this problem becomes linear when restricted to acyclic digraphs [12].

Among all digraphs, those which have the identity relation as the maximum bisimulation are of particular interest. For example, in set theory, these digraphs have been proposed by Aczel to describe hypersets [4, 13]. Observe that extensional acyclic digraphs are such digraphs, as shown in Lemma 1 below.

**Definition 5.** *A digraph* $D$ *is said to be* hyper-extensional *if every bisimulation over* $D$ *is contained in the identity relation, that is, if for every bisimulation* $B$ *over* $D$, *it holds that* $x = y$ *whenever* $xBy$.

It follows from the above-mentioned result of Paige and Tarjan that hyper-extensional digraphs can be recognized in time $O(|E(D)|\log|V(D)|)$. In general, a bisimulation is defined over a digraph $D$ whose arc relation $E(D)$ is an arbitrary binary relation (thus $D$ may have self-loops, or parallel arcs of opposite directions). However, the digraphs that will turn up in our NP-complete reduction are actually simple. The following lemma gives some properties of hyper-extensional simple digraphs.

**Lemma 1.** *Let* $D$ *be a hyper-extensional simple digraph. The following hold:*

i) $D$ *is extensional;*

ii) $D$ *has a sink;*

iii) *there is a directed path from every* $v \in V(D)$ *to a sink of* $D$;

iv) *every e.a. digraph is hyper-extensional.*

*Proof.* To see that $i)$ holds, it suffices to observe that if distinct $u, v \in V(D)$ have $N^+(u) = N^+(v)$, then the equivalence relation that puts $u$ and $v$ in the same class and keeps every other vertex in a singleton class, that is, the equivalence relation induced by the partition $\{\{u,v\}\} \cup \{\{w\} \mid w \in V(D) \setminus \{u,v\}\}$, is a non-trivial bisimulation over $D$.

If $ii)$ did not hold, then $D$ is due to have at least two vertices. Hence, the universal relation, that is, the equivalence relation that puts all vertices of $D$ in the same equivalence class, is a non-trivial bisimulation over $D$.

To show $iii)$, take, for a contradiction, a vertex $v \in V(D)$ so that there is no directed path from $v$ to a sink of $D$. Let $C$ be the set of all vertices $u$ such that there is a directed path from $v$ to $u$. By assumption on $v$, each vertex in $C$ has at least one out-neighbor, and all such out-neighbors are in $C$. Since $N^+(v) \neq \emptyset$, we have that $|C| \geq 2$. The equivalence relation induced by the partition $\{C\} \cup \{\{w\} \mid w \in V(D) \setminus C\}$ is a non-trivial bisimulation over $D$, contradicting the hyper-extensionality of $D$.

As far as $iv$) is concerned, suppose that $D$ is an e.a. digraph that admits a non-trivial bisimulation $B$, and let $x_0, y_0$ be two distinct vertices of $D$ so that $x_0 B y_0$. By the extensionality of $D$, $N^+(x_0) \neq N^+(y_0)$. Therefore, we may assume w.l.o.g. that there exists an $x_1 \in N^+(x_0) \setminus N^+(y_0)$. Since $x_0 B y_0$, there exists $y_1 \in N^+(y_0)$, thus $y_1 \neq x_1$, so that $x_1 B y_1$. We can repeat the above procedure indefinitely, and, as the number of vertices of $D$ is finite, we will reach a vertex $x_i$, or $y_i$, already visited. This contradicts the acyclicity of $D$. □

## 3. The complexity of set graph recognition and related problems

In this section, we prove that the following three problems are NP-complete:[1]

**Problem** EAO. *Given a graph $G$, decide whether $G$ is a set graph.*

**Problem** sEAO. *Given a graph $G$, decide whether $G$ admits a slim e.a.o.*

**Problem** HEO. *Given a graph $G$, decide whether $G$ admits a hyper-extensional orientation.*

Let HP denote the NP-complete HAMILTONIAN PATH problem [14]: Given a graph $G$, is there a Hamiltonian path in $G$, that is, a path meeting every vertex exactly once? To obtain the above results, we offer a reduction from the following variant of HP:

**Problem** HP′. *Given a graph $G$ with exactly two leaves, decide whether $G$ has a Hamiltonian path.*

To see that also Problem HP′ is NP-complete, the following reduction from Problem HP suffices. Given a graph $G$, construct $G^+$ having $V(G) \cup \{s_1, s_2, t_1, t_2\}$ as vertex set, and $E(G) \cup \{s_1 s_2, t_1 t_2\} \cup \{s_2 v, t_2 v \mid v \in V(G)\}$ as edge set. Clearly, $G$ has a Hamiltonian path if and only if $G^+$ has a Hamiltonian path (having $s_1$ and $t_1$ as endpoints). Moreover, the Hamiltonian paths of $G$ are in bijection with the Hamiltonian paths of $G^+$, an observation which will turn out useful in Section 4.

*3.1. Finding a (slim) extensional acyclic orientation*

Given a graph $G = (V, E)$, denote by $S(G)$ the *subdivision graph of $G$*, that is, the bipartite graph obtained by subdividing once every edge of $G$. Stated formally, $S(G) = (V \cup X, F)$, where

- $X = \{x^e \mid e \in E\}$
- $F = \{u x^{uv} \mid uv \in E\}$

A vertex of $X$ is called an *edge vertex*.

**Lemma 2.** *If $G$ is a graph with exactly two leaves that has a Hamiltonian path, then $S(G)$ admits a slim e.a.o.*

*Proof.* Let $(v_1, v_2, \ldots, v_n)$ be a Hamiltonian path in $G$. Then $s = v_1$ and $t = v_n$ are the two leaves of $G$. An edge vertex of $X$ is called *touched* if the above Hamiltonian path of $G$ uses the corresponding edge of $G$, and *untouched* otherwise. Partition $X$ as $X = T \cup U$ by distinguishing *touched* edge vertices from *untouched* ones. Choose any total order $\prec$ on the vertices of $S(G)$ with the following properties:

---

[1]The NP-completeness of set graph recognition was first announced at Bled '11 – 7th Slovenian International Conference on Graph Theory.

*i)* every vertex in $U$ is placed after any vertex in $T \cup V$;

  *ii)* $v_i \prec x^{v_i v_{i+1}} \prec v_{i+1}$, for every $i \in \{1, \ldots, n-1\}$.

Notice that such a total order exists. Consider the orientation $D$ of $S(G)$ such that every edge $uv \in E(S(G))$ is oriented in $D$ as $u \to v$ if and only if $u \succ v$. Clearly, this is an acyclic orientation. Furthermore, $D$ is also extensional, since:

- vertex $s = v_1$ is the only vertex with $N^+(s) = \emptyset$;

- every untouched vertex in $U$, say $x^{uv} \in U$, is the only vertex having $N^+(x^{uv}) = \{u, v\}$;

- every touched vertex in $T$, say $x^{v_i v_{i+1}} \in T$, is the only vertex having $N^+(x^{v_i v_{i+1}}) = \{v_i\}$;

- every vertex in $V \setminus \{s\}$, say $v_i \in V$ (with $2 \leq i \leq n$), is the only vertex with $N^+(v_i) = \{x^{v_{i-1} v_i}\}$.

To see that $D$ is also slim, observe first that the out-neighborhood of any vertex $v \in T \cup (V \setminus \{s\})$ is a singleton. Therefore, in the digraph obtained by removing the out-going arc from $v$, vertex $v$ collides with $s$. Finally, since both $s$ and $t$ are leaves in $G$, for every untouched vertex in $U$, say $x^{v_i v_j} \in U$, we have $i, j \in \{2, \ldots, n-1\}$. The removal of the arc $x^{v_i v_j} v_i$ creates a collision between $x^{v_i v_j}$ and $x^{v_j v_{j+1}}$, and similarly the removal of the arc $x^{v_i v_j} v_j$ creates a collision between $x^{v_i v_j}$ and $x^{v_i v_{i+1}}$. □

**Lemma 3.** *Let $G$ be a graph. If $S(G)$ admits an e.a.o., then $G$ has a Hamiltonian path.*

*Proof.* Let $D$ be an e.a.o. of $S(G)$ and let its sink be $v$. We claim that $D$ has a directed path passing through all the vertices of $G$, which hence produces a Hamiltonian path for $G$.

Indeed, let $P$ be a longest directed path in $D$ starting in a vertex of $G$ and ending at $v$. Let $u \in V(G)$ be the endpoint of $P$ other than $v$. If all vertices of $G$ are on $P$, we are done. If not, let $u'$ be a vertex of $G$ not on $P$. Let $Q$ be a longest directed path from $u'$ to $v$, and let $x$ be the first vertex on $Q$ that belongs to $P$. Let $y$ and $z$ ($y \neq z$) be the predecessors of $x$ on $P$ and on $Q$, respectively.

If $x$ is a vertex of $G$, then $y$ and $z$ are edge vertices (thus different from $u$ and $u'$). Note that by construction each of $y$ and $z$ have exactly two incident arcs, one in-coming, on $P$ or on $Q$, and one out-going to $x$. This implies that $N^+(y) = N^+(z) = \{x\}$, contradicting the extensionality of $D$.

Otherwise, $x$ is an edge vertex, and $x$ must be the sink of $D$, since its two incident arcs are in-coming. But $y$ and $z$ are again in collision, since from the maximality of the paths and the acyclicity of $D$ they cannot have other out-neighbors than $x$. □

**Theorem 1.** *Problems* EAO *and* sEAO *are* NP-*complete, even when the input is restricted to bipartite graphs with exactly two leaves.*

*Proof.* The problems belong to NP, since acyclicity, extensionality and slimness can be checked in polynomial time; actually, extensionality of an acyclic digraph can be verified in linear time [12]. The hardness follows by reducing from Problem HP$'$, by Lemmas 2 and 3. □

**Remark 1.** *Instead of requiring that the digraph obtained by removing any arc from an extensional acyclic digraph creates a collision (as in the definition of slimness), one can consider, in a similar way, extensional acyclic digraphs with the property that* reversing *any arc produces either a cycle or a collision. Notice that the slim e.a. orientation of $S(G)$ given in the proof of Lemma 2 has this property as well (in fact, reversing any arc produces a collision). In particular, this implies that it is* NP-*complete to verify whether a given bipartite graph with exactly two leaves admits an e.a.o. such that reversing any arc in it produces either a cycle or a collision.*

*3.2. Finding a hyper-extensional orientation*

Given digraphs $D_1$ and $D_2$ with disjoint vertex sets, and given vertices $v_i \in V(D_i)$, $i = 1, 2$, we denote by $U(D_1, v_1, v_2, D_2)$ the digraph obtained by taking a copy of $D_1$ and a copy of $D_2$ and adding the arc $v_1 \to v_2$. Formally, $U(D_1, v_1, v_2, D_2)$ has

- $V(D_1) \cup V(D_2)$ as vertex set,
- $E(D_1) \cup E(D_2) \cup \{v_1 \to v_2\}$ as the arc relation.

We define this operation analogously for graphs.

Our reduction will encode any graph $G$ having two leaves $s$ and $t$ by the graph $U(S(G), s, a_8, G_8)$, where $G_8$ is the underlying graph of digraph $D_8$, depicted in Figure 1. We start with a preliminary lemma.

**Lemma 4.** *Let $D_1$ and $D_2$ be two hyper-extensional digraphs. If the sink of $D_1$ is $s$ and if $D_2$ has a source $t$, then the digraph $U(D_1, s, t, D_2)$ is hyper-extensional.*

*Proof.* Let $D = U(D_1, s, t, D_2)$ and let $B$ be a bisimulation over $D$. Digraph $D$ is extensional, since $D_1$ and $D_2$ are extensional, by Lemma 1, and $t$ is a source of $D_2$. To prove that $x = y$ whenever $xBy$, we argue by contradiction, and consider three cases.

First, by construction, $B$ restricted to $V(D_2)$, that is, the relation $B_2 = \{(x, y) \mid xBy \land x, y \in V(D_2)\}$, is a bisimulation over $D_2$. Therefore, $xBy$ cannot hold for distinct $x, y \in V(D_2)$.

Second, suppose that $x_0 B y_0$ holds for (distinct) $x_0 \in V(D_1)$ and $y_0 \in V(D_2)$. Take $x_1 \in N^+(x_0)$ so that $x_1$ is a vertex on the directed path from $x_0$ to $s$ (or $x_1 = t$, if $x_0 = s$). Since $x_0 B y_0$, there exists $y_1 \in N^+(y_0)$, thus $y_1 \neq x_1$, such that $x_1 B y_1$. By repeating the above procedure sufficiently many times, we reach a pair $(x_i, y_i)$ (where $i \geq 0$) such that $x_i = s$, $y_i \in V(D_2)$ and $sBy_i$. Since $N^+(s) = \{t\}$, there exists a $y_{i+1} \in N^+(y_i)$ so that $tBy_{i+1}$. Recall that $t$ is a source of $D_2$, therefore $t \neq y_{i+1}$. This contradicts the previous case.

Finally, we claim that also the restriction of $B$ to $V(D_1)$, that is the relation $B_1 = \{(x, y) \mid xBy \land x, y \in V(D_1)\}$, is a bisimulation over $D_1$. Observe that neither $sBx$ nor $xBs$ can hold for $x \in V(D_1) \setminus \{s\}$. This is true, since $N^+(s) = \{t\}$, and, by the previous case, there can be no $x_1 \in V(D_1)$ such that $tBx_1$, or $x_1Bt$. Hence, also $sB_1x$ or $xB_1s$ cannot hold for $x \in V(D_1) \setminus \{s\}$. If $xB_1y$, for distinct $x, y \in V(D_1)$ and $s \notin \{x, y\}$, then both conditions *i*) and *ii*) of the bisimulation definition hold, by construction and by the fact that $xBy$. Therefore, $B_1$ is a bisimulation over $D_1$, and by hyper-extensionality of $D_1$ it follows that $x = y$, a contradiction. □

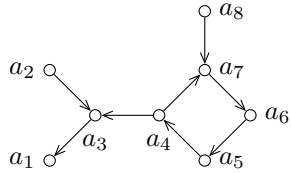

Figure 1: Digraph $D_8$ (denote by $G_8$ its underlying graph); $D_8$ is a gadget to force a sink when the orientation can have cycles; one of $a_1$ or $a_2$ must be a sink in any *extensional* orientation of $G_8$.

**Lemma 5.** *If $G$ is a graph with two leaves $s$ and $t$, and if $G$ has a Hamiltonian path, then the graph $U(S(G), s, a_8, G_8)$ admits a hyper-extensional orientation.*

*Proof.* First, let $D$ be the e.a.o. of $S(G)$ obtained as explained in the proof of Lemma 2, where $s$ is taken to be its sink. By Lemma 1, $D$ is also hyper-extensional. Next, observe that also $D_8$ is hyper-extensional, by applying, for example, the partition refinement algorithm of [11]. Since $a_8$ is a source of $D_8$, by Lemma 4, the digraph $U(D, s, a_8, D_8)$ is hyper-extensional, which proves the claim. □

**Lemma 6.** *Let $G$ be a graph. If $U(S(G), s, a_8, G_8)$ admits a hyper-extensional orientation, then $G$ has a Hamiltonian path.*

*Proof.* We reason as in the proof of Lemma 3. Let $D$ be a hyper-extensional orientation of $U(S(G), s, a_8, G_8)$. Therefore, $D$

- is extensional,
- has a (unique) sink $v$ that belongs to $G_8$, and
- from every vertex of $D$ there is a directed path to $v$.

We claim that $D$ has a directed path passing through all the vertices of $G$, which thus produces a Hamiltonian path for $G$.

Indeed, let $P$ be a longest directed path in $D$ starting in a vertex of $G$ and ending at $v$, the sink of $D$. Let $u \in V(G)$ be the endpoint of $P$ other than $v$. If all vertices of $G$ are on $P$, we are done. If not, let $u'$ be a vertex of $G$ not on $P$. Let $Q$ be a longest directed path from $u'$ to $v$, and let $x$ be the first vertex on $Q$ that belongs to $P$. From construction, we have that $x$ is a vertex of $S(G)$. Let $y$ and $z$ ($y \neq z$) be the predecessors of $x$ on $P$ and on $Q$, respectively.

If $x$ is a vertex of $G$, then $y$ and $z$ are edge vertices (thus different from $u$ and $u'$). Note that by construction, each of $y$ and $z$ have exactly two incident arcs, one in-coming, on $P$ or on $Q$, and one out-going to $x$. This implies that $N^+(y) = N^+(z) = \{x\}$, contradicting the extensionality of $D$.

Otherwise, $x$ is an edge vertex, and $x$ must be the sink of $D$, since its two incident arcs are in-coming. This contradicts the fact that the sink of $D$ is a vertex of $G_8$. □

**Theorem 2.** *Problem HEO is NP-complete, even when the input is restricted to bipartite graphs with exactly three leaves.*

*Proof.* The problem belongs to NP, since hyper-extensionality can by checked in polynomial time, for example by the algorithm of [11]. The hardness follows by reducing from Problem HP′, by Lemmas 5 and 6. □

## 4. The complexity of counting extensional orientations

We denote by #EAO, #sEAO, #HEO, #HP and #HP′ the corresponding counting variants of the problems considered in Section 3. For instance, in the #EAO problem the task is to determine the number of all e.a.o.s of a given graph.

Since Problem #HP is #P-complete (see [15, Ch.18],[16]), the simple reduction we gave at the beginning of Section 3 implies that Problem #HP′ is #P-complete as well.

**Theorem 3.** *Problems #EAO and #sEAO are #P-complete, even when the input is restricted to bipartite graphs with exactly two leaves.*

*Proof.* We first show that if $G$ is a graph with two leaves, $s$ and $t$, then any e.a.o. $D$ of $S(G)$ is slim. As argued in the proof of Lemma 3, the vertices of $G$ belong to a directed path $P$ of $D$. The endpoints of $P$ are $s$ and $t$, since they are leaves in $G$. We may assume that $s$ is its last vertex, so that $s$ is the sink of $D$. Denote by $(t = v_n, v_{n-1}, \ldots, v_1 = s)$ the order in which the vertices of $G$ appear on $P$. By the construction of $S(G)$ and the fact that $P$ is a directed path, $N^+(x^{v_{i+1}v_i}) = \{v_i\}$, for every $1 \leqslant i \leqslant n-1$. Every vertex of $D$ not on $P$ is an edge vertex $x^{v_iv_j}$, with $i, j \in \{2, \ldots, n-1\}$. If $x^{v_iv_j}$ had only one out-neighbor in $D$, say $v_i$, then it would be in collision with $x^{v_{i+1}v_i}$. Therefore, $N^+(x^{v_iv_j}) = \{v_i, v_j\}$. We can conclude that $N^+(v_i) = \{x^{v_iv_{i-1}}\}$, for every $2 \leqslant i \leqslant n$. This shows that $D$ is an orientation obtained as explained in Lemma 2, hence it is also slim.

We now reduce from #HP′. If $G$ is a graph with two leaves $s$ and $t$, every Hamiltonian path in $G$ between $s$ and $t$ induces two slim e.a.o.s (having either $s$ or $t$ as sink) for $S(G)$, as argued in Lemma 2. Moreover, different Hamiltonian paths of $G$ induce different pairs of such slim e.a.o.s for $S(G)$. Conversely, by Lemma 3 and the above argument, every e.a.o. of $S(G)$ is slim, and it induces a Hamiltonian path of $G$. This shows that the number of (slim) e.a.o.s of $S(G)$ is exactly twice the number of Hamiltonian paths in $G$, hence Problems #EAO and #sEAO are #P-complete. □

**Remark 2.** *The above proof implies that it is #P-complete to determine the number of all e.a. orientations in which any arc reversal produces either a cycle or a collision, even when the input is restricted to bipartite graphs with exactly two leaves.*

Regarding Problem #HEO, we need the following lemma, describing the possible orientations of the graph $G_8$.

**Lemma 7.** *The digraph $D_8$ depicted in Figure 1, together with $D_8'$, the digraph obtained from $D_8$ by reversing the arcs $a_2a_3$ and $a_3a_1$, are the only digraphs having $G_8$ as underlying graph and satisfying properties i), ii) and iii) stated in Lemma 1.*

*Proof.* Suppose for a contradiction that there exists an orientation $D$ of $G_8$ different from $D_8$ or $D_8'$ and satisfying the properties $i)$, $ii)$ and $iii)$ stated in Lemma 1.

Note that exactly one of $a_1$ or $a_2$ must be a sink in $D$, as otherwise they would have the same out-neighborhood. Say $a_3 \to a_1$ and $a_2 \to a_3$. This implies that $a_8 \to a_7$. Since from every vertex of $D$ there is a directed path to $a_1$, we have $a_4 \to a_3$. Since $N^+(a_2) \neq N^+(a_4)$, we have that $a_4$ has at least one other out-neighbor.

Assume first that $a_7 \to a_4$, and hence $a_4 \to a_5$. Since $N^+(a_5) \neq \emptyset$, we have $a_5 \to a_6$. Similarly, $a_6 \to a_7$. Hence, $N^+(a_6) = N^+(a_8)$, contradicting the extensionality of $D$. Otherwise, since from $a_7$ there must be a directed path to $a_1$, we have $a_7 \to a_6 \to a_5 \to a_4$, which contradicts the fact that $D$ is not $D_8$, nor $D_8'$. □

**Theorem 4.** *Problem #HEO is #P-complete, even when the input is restricted to bipartite graphs with exactly three leaves.*

*Proof.* We reduce again from #HP′. If $G$ is a graph with two leaves $s$ and $t$, every Hamiltonian path in $G$ between $s$ and $t$ induces a pair of hyper-extensional orientations of $U(S(G), s, a_8, G_8)$. Indeed, the edges between vertices of $S(G)$ can be oriented as in Lemma 2 (taking $s$ as a 'local' sink for $S(G)$), whereas the edges between the vertices of $G_8$ can be oriented as in $D_8$ or as in $D_8'$. Moreover, different Hamiltonian paths of $G$ induce different pairs of such hyper-extensional orientations of $U(S(G), s, a_8, G_8)$.

Conversely, if $D$ is a hyper-extensional orientation of $U(S(G), s, a_8, G_8)$, Lemma 6 and the argument employed in the proof of Theorem 3 show that $D[V(S(G))]$ must be oriented as indicated by the proof of Lemma 2. Moreover, Lemma 7 shows that $D[V(G_8)]$ is either $D_8$ or $D_8'$. This allows us to conclude that the number of hyper-extensional orientations of $U(S(G), s, a_8, G_8)$ is exactly twice the number of Hamiltonian paths in $G$. Hence also Problem #HEO is #P-complete. □

## 5. The complexity of finding a separating code for extensional digraphs

Given a vertex $v$ in a graph $G$, denote by $N[v]$ the *closed* neighborhood of $v$, that is $N(v) \cup \{v\}$. Similarly, for a digraph $D$ and a vertex $v \in V(D)$, the closed in-neighborhood of $v$ is $N^-[v] = N^-(v) \cup \{v\}$. Given a graph $G$, a subset $C \subseteq V(G)$ is called:

- *dominating set*, if for all $v \in V(G)$, $N[v] \cap C \neq \emptyset$, cf. [17];
- *separating code*, if for distinct $u, v \in V(G)$ it holds $N[u] \cap C \neq N[v] \cap C$, cf. [18];
- *identifying code*, if $C$ is a dominating set and a separating code, cf. [19];

Moreover, if $G$ is a bipartite graph $G = (A \cup B, E)$ (with no edges within $A$ or within $B$), then a set $C \subseteq B$ is called

- *discriminating code*, if for all $v \in A$, $N(v) \cap C \neq \emptyset$ and for distinct $u, v \in A$ it holds $N(u) \cap C \neq N(v) \cap C$, [20, 21].

In case of digraphs, these notions have been analogously defined in terms of *in*-neighbors. Given a digraph $D$, a subset $C \subseteq V(D)$ is called:

- *dominating set*, if for all $v \in V(G)$, $N^-[v] \cap C \neq \emptyset$, cf. [17];
- *separating code*, if for distinct $u, v \in V(G)$ it holds $N^-[u] \cap C \neq N^-[v] \cap C$ [18];
- *identifying code*, if $C$ is a dominating set and a separating code [22];

In this section, we are concerned with separating codes in digraphs, with two minor changes: we will be referring to *(open) out*-neighborhoods, instead of closed in-neighborhoods:

**Definition 6.** *Given a digraph $D$ and $C \subseteq V(D)$ we say that $C$ is an* open-out-separating code *if for distinct $u, v \in V(G)$ it holds $N^+(u) \cap C \neq N^+(v) \cap C$.*

It can be easily seen that a digraph $D$ has an open-out-separating code if and only if $D$ is extensional.

The problem of finding the minimum size of a separating code of a given graph was shown to be NP-complete in [23, 24]. An analogous result holds for digraphs [22], even when restricted to acyclic instances. In what follows, we will show that finding the minimum size of an open-out-separating code is NP-complete, even when restricted to (extensional) acyclic digraphs.

**Problem ooSC.** *Given a digraph $D$ and an integer $k$, decide whether $D$ has an open-out-separating code $C$ of size at most $k$.*

The following problem was shown to be NP-complete in [21].

**Problem DC.** *Given a bipartite graph $G = (A \cup B, E)$ and an integer $k$, decide whether there exists a discriminating code $C \subseteq B$ of size at most $k$.*

**Theorem 5.** *Problem ooSC is NP-complete, even when the input is restricted to extensional acyclic digraphs.*

*Proof.* Reduce from Problem DC. Let $G = (A \cup B, E)$ be a bipartite graph (with no edges within $A$ or within $B$), where $B = \{b_1, \ldots, b_m\}$, $m \geqslant 1$. Construct the acyclic digraph $D = (V, F)$ as follows:

- $V = A \cup B \cup \{c_0, c_1, \ldots, c_m\}$,
- $F = \{a \to b \mid a \in A \wedge b \in B \wedge ab \in E\} \cup \{b_i \to c_i \mid 1 \leqslant i \leqslant m\} \cup \{c_i \to c_j \mid 1 \leqslant i \leqslant m, 0 \leqslant j < i\}$.

We claim that $G$ has a discriminating code of size at most $k$ if and only if $D$ has an open-out-separating code of size at most $k + m + 1$.

For the forward implication, note that if $C$ is a discriminating code for $G$, then $C \cup \{c_0, \ldots, c_m\}$ is an open-out-separating code for $D$.

For the reverse implication, let $C$ be an open-out-separating code for $D$. We show that $c_0, \ldots, c_m \in C$. First, $c_0 \in C$, as otherwise $N(c_1) \cap C = \emptyset = N^+(c_0) = N^+(c_0) \cap C$. Assuming now that $c_0, \ldots, c_i$, $0 \leqslant i \leqslant m - 2$, belong to $C$, note that $c_{i+1} \in C$ as well, as otherwise $N^+(c_{i+2}) \cap C = N^+(c_{i+1}) \cap C$. Therefore, $c_0, \ldots, c_{m-1} \in C$. Additionally, $c_m \in C$, as otherwise $N^+(b_m) \cap C = \emptyset = N^+(c_0) \cap C$. This concludes the proof, since $C \cap B$ is a discriminating code for $G$.

Note that if for distinct $a_1, a_2 \in A$, $N_G(a_1) \neq N_G(a_2)$ holds (which can be assumed w.l.o.g., since otherwise $G$ has no discriminating code), then $D$ is also extensional. □

## 6. Conclusions

In this paper we continued the study of set graphs, and provided several NP-completeness and #P-completeness results related to finding and counting extensional acyclic, slim extensional acyclic, and hyper-extensional orientations. In particular, exploiting connections between extensional acyclic orientations and Hamiltonian paths in certain graphs, we showed that problems EAO, sEAO and HEO are NP-complete, and their counting variants #P-complete, even for restricted bipartite graphs. We also showed that it is NP-complete to determine whether an extensional acyclic digraph contains an open-out-separating code of given size.

Let us conclude the paper with mentioning some open questions and possibilities for further research in this area. In [2], it was showed that every connected claw-free graph is a set graph, and a polynomial time algorithm was given for finding an extensional acyclic orientation of a claw-free set graph. Hence, in view of the fact that the EAO problem is (trivially) polynomial for claw-free graphs, it is interesting to study the complexity of the problem for various generalizations of claw-free graphs. For example, the EAO problem is NP-complete for the class of graphs in which no two induced claws have an edge in common.[2] A more restricted class but still a generalization of the class of claw-free graphs is the class of claw disjoint graphs [25]: A graph is said to be *claw disjoint* if no two induced claws in it have a vertex in common. Further examples of generalizations of the class of claw-free graphs can be obtained by forbidding, as an induced subgraph, some subdivision of the claw. For example, a *fork* is the graph obtained from a claw by subdividing one of its edges. It remains an open problem to determine the complexity of the EAO problem for claw disjoint graphs, as well as for fork-free graphs and more generally for $F$-free graphs, where $F$ is any proper subdivision of the claw.

---

[2]To see this, observe that the following simplified version of Lemma 2 holds: "If $G$ is a graph with a Hamiltonian path, then $S(G)$ admits an e.a.o." Since the Hamiltonian path problem remains NP-complete for cubic graphs [14], we can transform a given cubic graph $G$ to its subdivision graph $S(G)$, which is clearly a graph in which no two claws have an edge in common. Together with Lemma 3, this establishes the claim.

Another related and yet unsettled question is that of determining the complexity of the #EAO problem for claw-free graphs.

Finally, since the EAO problem is NP-complete for bipartite graphs but polynomial for trees, it would be interesting to determine the complexity of this problem for other classes of perfect graphs, such as threshold graphs, split graphs, cographs, interval graphs, chordal graphs.

**Acknowledgements**

A.T. is grateful to Alberto Policriti for his encouragement and fruitful discussions which played a key role in getting this research started. M.M. is supported in part by "Agencija za raziskovalno dejavnost Republike Slovenije", research program P1-0285 and research projects J1-4010 and J1-4021.